# The Complexity of Online Manipulation of Sequential Elections

Edith Hemaspaandra
Dept. of Computer Science
Rochester Inst. of Technology
Rochester, NY 14623, USA
www.cs.rit.edu/~eh

Lane A. Hemaspaandra
Dept. of Computer Science
University of Rochester
Rochester, NY 14627, USA
www.cs.rochester.edu/u/lane

Jörg Rothe
Institut für Informatik
Universität Düsseldorf
40225 Düsseldorf, Germany
rothe@cs.uni-duesseldorf.de

## ABSTRACT

Most work on manipulation assumes that all preferences are known to the manipulators. However, in many settings elections are open and sequential, and manipulators may know the already cast votes but may not know the future votes. We introduce a framework, in which manipulators can see the past votes but not the future ones, to model online coalitional manipulation of sequential elections, and we show that in this setting manipulation can be extremely complex even for election systems with simple winner problems. Yet we also show that for some of the most important election systems such manipulation is simple in certain settings. This suggests that when using sequential voting, one should pay great attention to the details of the setting in choosing one's voting rule.

Among the highlights of our classifications are: We show that, depending on the size of the manipulative coalition, the online manipulation problem can be complete for each level of the polynomial hierarchy or even for PSPACE. We obtain the most dramatic contrast to date between the nonunique-winner and unique-winner models: Online weighted manipulation for plurality is in P in the nonunique-winner model, yet is coNP-hard (constructive case) and NP-hard (destructive case) in the unique-winner model. And we obtain what to the best of our knowledge are the first $P^{NP[1]}$-completeness and $P^{NP}$-completeness results in the field of computational social choice, in particular proving such completeness for, respectively, the complexity of 3-candidate and 4-candidate (and unlimited-candidate) online weighted coalition manipulation of veto elections.

## Categories and Subject Descriptors

I.2.11 [**Artificial Intelligence**]: Distributed Artificial Intelligence—*Multiagent systems*; F.1.2 [**Computation by Abstract Devices**]: Modes of Computation; F.2.2 [**Analysis of Algorithms and Problem Complexity**]: Nonnumerical Algorithms and Problems

## General Terms

Theory

## Keywords

Computational complexity, computational social choice,

elections, manipulation, online algorithms, preferences, sequential voting

## 1. INTRODUCTION

Voting is a widely used method for preference aggregation and decision-making. In particular, *strategic* voting (or *manipulation*) has been studied intensely in social choice theory (starting with the celebrated work of Gibbard [19] and Satterthwaite [29]) and, in the rapidly emerging area of *computational* social choice, also with respect to its algorithmic properties and computational complexity (starting with the seminal work of Bartholdi, Tovey, and Trick [3]; see the surveys [15, 16]). This computational aspect is particularly important in light of the many applications of voting in computer science, ranging from meta-search heuristics for the internet [14], to recommender systems [18] and multiagent systems in artificial intelligence (see the survey by Conitzer [11]).

Most of the previous work on manipulation, however, is concerned with voting where the manipulators know the nonmanipulative votes. Far less attention has been paid (see the related work below) to manipulation in the midst of elections that are modeled as dynamic processes.

We introduce a novel framework for online manipulation, where voters vote in sequence and the current manipulator, who knows the previous votes and which voters are still to come but does not know their votes, must decide—right at that moment—what the "best" vote to cast is. So, while other approaches to sequential voting are stochastic, game-theoretic (yet different from our approach, see Footnote 1), or axiomatic in nature (again, see the related work), our approach to manipulation of sequential voting is shaped by the area of "online algorithms" [8], in the technical sense of a setting in which one (for us, each manipulative voter) is being asked to make a manipulation decision just on the basis of the information one has in one's hands at the moment even though additional information/system evolution may well be happening down the line. In this area, there are different frameworks for evaluation. But the most attractive one, which pervades the area as a general theme, is the idea that one may want to "maxi-min" things—*one may want to take the action that maximizes the goodness of the set of outcomes that one can expect regardless of what happens down the line from one time-wise*. For example, if the current manipulator's preferences are Alice > Ted > Carol > Bob and if she can cast a (perhaps insincere) vote that ensures that Alice or Ted will be a winner no matter what later voters do, and there is no vote she can cast that ensures that Alice will always be a winner, this maxi-min approach would say that that vote is a "best" vote to cast.

It will perhaps be a bit surprising to those familiar with online algorithms and competitive analysis that in our model of online manipulation we will not use a (competitive) *ratio*. The reason is that voting commonly uses an *ordinal* preference model, in which preferences are total orders of the candidates. It would be a severely improper step to jump from that to assumptions about intensity of preferences and utility, e.g., to assuming that everyone likes her $n$th-to-least favorite candidate exactly $n$ times more than she likes her least favorite candidate.

*Related Work.*

Conitzer and Xia [37] (see also the related paper by Desmedt and Elkind [13]) define and study the Stackelberg voting game (also quite naturally called, in an earlier paper that mostly looked at two candidates, the roll-call voting game [30]). This basically is an election in which the voters vote in order, *and the preferences are common knowledge—everyone knows everyone else's preferences, everyone knows that everyone knows everyone else's preferences, and so on out to infinity*. Their analysis of this game is game-theoretically shaped; they compute a subgame perfect Nash equilibrium from the back end forward. Under their work's setting and assumptions, for bounded numbers of manipulators manipulation is in P, but we will show that in our model even with bounded numbers of manipulators manipulation sometimes (unless P = NP) falls beyond P.[1]

The interesting "dynamic voting" work of Tennenholtz [33] investigates sequential voting, but focuses on axioms and voting rules rather than on coalitions and manipulation. Much heavily Markovian work studies sequential decision-making and/or dynamically varying preferences; our work in contrast is nonprobabilistic and focused on the complexity of coalitional manipulation. Also somewhat related to, but quite different from, our work is the work on possible and necessary winners. The seminal paper on that is due to Konczak and Lang [25], and more recent work includes [36, 7, 1, 5, 6, 10, 4, 27]; the biggest difference is that those are, loosely, one-quantifier settings, but the more dynamic setting of online manipulation involves numbers of quantifiers that can grow with the input size. Another related research line studies multi-issue elections [38, 39, 40, 41]; although there the separate issues may run in sequence, each issue typically is voted on simultaneously and with preferences being common knowledge.

## 2. PRELIMINARIES

*Elections.*

A *(standard, i.e., simultaneous) election* $(C, V)$ is specified by a set $C$ of candidates and a list $V$, where we assume that each element in $V$ is a pair $(v, p)$ such that $v$ is a voter name and $p$ is $v$'s vote. How the votes in $V$ are represented depends on the election system used—we assume, as is required by most systems, votes to be total preference orders over $C$. For example, if $C = \{a, b, c\}$, a vote of the form $c > a > b$ means that this voter (strictly) prefers $c$ to $a$ and $a$ to $b$.

We introduce election snapshots to capture sequential election scenarios as follows. Let $C$ be a set of candidates and let $u$ be (the name of) a voter. An *election snapshot for $C$ and $u$* is specified by a triple $V = (V_{<u}, u, V_{u<})$ consisting of all voters in the order they vote, along with, for each voter before $u$ (i.e., those in $V_{<u}$), the vote she cast, and for each voter after $u$ (i.e., those in $V_{u<}$), a bit specifying if she is part of the manipulative coalition (to which $u$ always belongs). That is, $V_{<u} = ((v_1, p_1), (v_2, p_2), \ldots, (v_{i-1}, p_{i-1}))$, where the voters named $v_1, v_2, \ldots, v_{i-1}$ (including perhaps manipulators and nonmanipulators) have already cast their votes (preference order $p_j$ being cast by $v_j$), and $V_{u<} = ((v_{i+1}, x_{i+1}), (v_{i+2}, x_{i+2}), \ldots, (v_n, x_n))$ lists the names of the voters still to cast their votes, in that order, and where $x_j = 1$ if $v_j$ belongs to the manipulative coalition and $x_j = 0$ otherwise.

*Scoring Rules.*

A *scoring rule* for $m$ candidates is given by a scoring vector $\alpha = (\alpha_1, \alpha_2, \ldots, \alpha_m)$ of nonnegative integers such that $\alpha_1 \geq \alpha_2 \geq \cdots \geq \alpha_m$. For an election $(C, V)$, each candidate $c \in C$ scores $\alpha_i$ points for each vote that ranks $c$ in the $i$th position. Let $score(c)$ be the total score of $c \in C$. All candidates scoring the most points are winners of $(C, V)$. Some of the most popular voting systems are *k-approval* (especially *plurality*, aka 1-approval) and *k-veto* (especially *veto*, aka 1-veto). Their $m$-candidate, $m \geq k$, versions are defined by the scoring vectors $(\underbrace{1, \ldots, 1}_{k}, \underbrace{0, \ldots, 0}_{m-k})$ and $(\underbrace{1, \ldots, 1}_{m-k}, \underbrace{0, \ldots, 0}_{k})$.
When $m$ is not fixed, we omit the phrase "$m$-candidate."

*Manipulation.*

The *(standard) weighted coalitional manipulation problem* [12], $\mathcal{E}$-Weighted-Coalitional-Manipulation (abbreviated by $\mathcal{E}$-WCM), for any election system $\mathcal{E}$ is defined as follows: Given a candidate set $C$, a list $S$ of nonmanipulative voters each having a nonnegative integer weight, a list $T$ of the nonnegative integer weights of the manipulative voters (whose preferences over $C$ are unspecified), with $S \cap T = \emptyset$, and a distinguished candidate $c \in C$, can the manipulative votes $T$ be set such that $c$ is a (or the) $\mathcal{E}$ winner of $(C, S \cup T)$?

Asking whether $c$ can be made "a winner" is called the nonunique-winner model and is the model of all notions in this paper unless mentioned otherwise. If one asks whether $c$ can be made a "one and only winner," that is called the unique-winner model. We also use the *unweighted* variant, where each vote has unit weight, and write $\mathcal{E}$-UCM as a shorthand. Note that $\mathcal{E}$-UCM with a *single* manipulator (i.e., $\|T\| = 1$ in the problem instance) is the manipulation problem originally studied in [3, 2]. Conitzer, Sandholm, and Lang [12] also introduced the *destructive* variants of these manipulation problems, where the goal is not to make $c$ win but to ensure that $c$ is not a winner, and we denote the corresponding problems by $\mathcal{E}$-DWCM and $\mathcal{E}$-DUCM. Finally, we write $\mathcal{E}$-WC$_{\neq\emptyset}$M, $\mathcal{E}$-UC$_{\neq\emptyset}$M, $\mathcal{E}$-DWC$_{\neq\emptyset}$M, and $\mathcal{E}$-DUC$_{\neq\emptyset}$M to indicate that the problem instances are required to have a nonempty coalition of manipulators.

---

[1] Our work too is game-theoretically connected. Although in our model we are asking whether we can reach our goal no matter what the future nonmanipulators do, if one thinks about what the actual effect of this is, one can see that our setting is in effect well-captured by what is known as a 2-player combinatorial game (combinatorial games are a particular type of complete-information sequential game). In our setting, the goal of one player in this game will be to ensure that the winner set (which of course heavily depends on what moves have occurred already and on the election system) will have nonempty intersection with a certain subset of the candidates, and the goal of the other player will be to ensure that that does not happen. Of course, the former player is in effect the currently-under-consideration and still-to-vote members of the manipulative coalition, and the latter player is capturing the same except regarding nonmanipulators. So, the key differences between [37] and our work regard goals and coalitionality. For them, each player (and they may have many players) is in effect a completely separate agent, with a preference order, and is trying to see if a change as an individual will make a more preferred candidate win. For us, the manipulative voters function as a coalition, and one that has an all-or-nothing goal, and there are no gradations within that goal in terms of our analysis (despite the fact that we use a preference order when speaking of the coalition), and we are in effect a two-player combinatorial game.

*Complexity-Theoretic Background.*

We assume the reader is familiar with basic complexity-theoretic notions such as the complexity classes P and NP, the class FP of polynomial-time computable functions, polynomial-time many-one reducibility ($\leq_m^p$), and hardness and completeness with respect to $\leq_m^p$ for a complexity class.

Meyer and Stockmeyer [28] and Stockmeyer [31] introduced and studied the polynomial hierarchy, PH $= \bigcup_{k \geq 0} \Sigma_k^p$, whose levels are inductively defined by $\Sigma_0^p = $ P and $\Sigma_{k+1}^p =$ NP$^{\Sigma_k^p}$, and their co-classes, $\Pi_k^p = $ co$\Sigma_k^p$ for $k \geq 0$. They also characterized these levels by polynomially length-bounded alternating existential and universal quantifiers. P$^{\text{NP}}$ is the class of problems solvable in deterministic polynomial time with access to an NP oracle. P$^{\text{NP}[1]}$ is the restriction of P$^{\text{NP}}$ where only one oracle query is allowed. P $\subseteq$ NP $\cap$ coNP $\subseteq$ NP $\cup$ coNP $\subseteq$ P$^{\text{NP}[1]}$ $\subseteq$ P$^{\text{NP}}$ $\subseteq$ $\Sigma_2^p \cap \Pi_2^p \subseteq \Sigma_2^p \cup \Pi_2^p \subseteq$ PH $\subseteq$ PSPACE, where PSPACE is the class of problems solvable in polynomial space. The *quantified boolean formula problem*, QBF, is a standard PSPACE-complete problem. QBF$_k$ ($\widetilde{\text{QBF}}_k$) denotes the restriction of QBF with at most $k$ quantifiers that start with $\exists$ ($\forall$) and then alternate between $\exists$ and $\forall$, and we assume that each $\exists$ and $\forall$ quantifies over a set of boolean variables. For each $k \geq 1$, QBF$_k$ is $\Sigma_k^p$-complete and $\widetilde{\text{QBF}}_k$ is $\Pi_k^p$-complete [32, 35].

## 3. OUR MODEL OF ONLINE MANIPULATION

The core of our model of online manipulation in sequential voting is what we call the *magnifying-glass moment*, namely, the moment at which a manipulator $u$ is the one who is going to vote, is aware of what has happened so far in the election (and which voters are still to come, but in general not knowing what they want, except in the case of voters, if any, who are coalitionally linked to $u$). In this moment, $u$ seeks to "figure out" what the "best" vote to cast is. We will call the information available in such a moment an *online manipulation setting* (*OMS*, for short) and define it formally as a tuple $(C, u, V, \sigma, d)$, where $C$ is a set of candidates; $u$ is a distinguished voter; $V = (V_{<u}, u, V_{u<})$ is an election snapshot for $C$ and $u$; $\sigma$ is the preference order of the manipulative coalition to which $u$ belongs; and $d \in C$ is a distinguished candidate. Given an election system $\mathcal{E}$, define the problem online-$\mathcal{E}$-Unweighted-Coalitional-Manipulation (abbreviated by online-$\mathcal{E}$-UCM), as follows: Given an OMS $(C, u, V, \sigma, d)$ as described above, does there exist some vote that $u$ can cast (assuming support from the manipulators coming after $u$) such that no matter what votes are cast by the nonmanipulators coming after $u$, there exists some $c \in C$ such that $c \geq_\sigma d$ and $c$ is an $\mathcal{E}$ winner of the election? By "support from the manipulators coming after $u$" we mean that $u$'s coalition partners coming after $u$, when they get to vote, will use their then-in-hand knowledge of all votes up to then to help $u$ reach her goal: By a joint effort $u$'s coalition can ensure that the $\mathcal{E}$ winner set will always include a candidate liked by the coalition as much as or more than $d$, even when the nonmanipulators take their strongest action so as to prevent this. Note that this candidate, $c$ in the problem description, may be different based on the nonmanipulators' actions. (Nonsequential manipulation problems usually focus on whether a single candidate can be made to win, but in our setting, this "that person or better" focus is more natural.) For the case of weighted manipulation, each voter also comes with a nonnegative integer weight. We denote this problem by online-$\mathcal{E}$-WCM.

We write online-$\mathcal{E}$-UCM[$k$] in the unweighted case and online-$\mathcal{E}$-WCM[$k$] in the weighted case to denote the problem when the number of manipulators from $u$ onward is restricted to be at most $k$.

Denote the corresponding destructive problems by online-$\mathcal{E}$-DUCM, online-$\mathcal{E}$-DWCM, online-$\mathcal{E}$-DUCM[$k$], and online-$\mathcal{E}$-DWCM[$k$]. In online-$\mathcal{E}$-DUCM we ask whether the given current manipulator $u$ (assuming support from the manipulators after her) can cast a vote such that no matter what votes are cast by the nonmanipulators after $u$, no $c \in C$ with $d \geq_\sigma c$ is an $\mathcal{E}$ winner of the election, i.e., $u$'s coalition can ensure that the $\mathcal{E}$ winner set never includes $d$ or any even more hated candidate. The other three problems are defined analogously.

Note that online-$\mathcal{E}$-UCM generalizes the original unweighted manipulation problem with a single manipulator as introduced by Bartholdi, Tovey, and Trick [3]. Indeed, their manipulation problem in effect is the special case of online-$\mathcal{E}$-UCM when restricted to instances where there is just one manipulator, she is the last voter to cast a vote, and $d$ is the coalition's most preferred candidate. Similarly, online-$\mathcal{E}$-WCM generalizes the (standard) coalitional weighted manipulation problem (for nonempty coalitions of manipulators). Indeed, that traditional manipulation problem is the special case of online-$\mathcal{E}$-WCM, restricted to instances where only manipulators come after $u$ and $d$ is the coalition's most preferred candidate. If we take an analogous approach except with $d$ restricted now to being the most hated candidate of the coalition, we generalize the corresponding notions for the destructive cases. We summarize these observations as follows.

PROPOSITION 1. *For each election system $\mathcal{E}$, it holds that (1) $\mathcal{E}$-UC$_{\neq\emptyset}$M $\leq_m^p$ online-$\mathcal{E}$-UCM, (2) $\mathcal{E}$-WC$_{\neq\emptyset}$M $\leq_m^p$ online-$\mathcal{E}$-WCM, (3) $\mathcal{E}$-DUC$_{\neq\emptyset}$M $\leq_m^p$ online-$\mathcal{E}$-DUCM, and (4) $\mathcal{E}$-DWC$_{\neq\emptyset}$M $\leq_m^p$ online-$\mathcal{E}$-DWCM.*

Corollary 2 below follows immediately.

COROLLARY 2. *(1) For each election system $\mathcal{E}$ such that the (unweighted) winner problem is solvable in polynomial time, it holds that $\mathcal{E}$-UCM $\leq_m^p$ online-$\mathcal{E}$-UCM. (2) For each election system $\mathcal{E}$ such that the weighted winner problem is solvable in polynomial time, it holds that $\mathcal{E}$-WCM $\leq_m^p$ online-$\mathcal{E}$-WCM. (3) For each election system $\mathcal{E}$ such that* *the winner problem is solvable in polynomial time, it holds that $\mathcal{E}$-DUCM $\leq_m^p$ online-$\mathcal{E}$-DUCM. (4) For each election system $\mathcal{E}$ such that the weighted winner problem is solvable in polynomial time, it holds that $\mathcal{E}$-DWCM $\leq_m^p$ online-$\mathcal{E}$-DWCM.*

We said above that, by default, we will use the *nonunique-winner model* and all the above problems are defined in this model. However, we will also have some results in the *unique-winner model*, which will, here, sharply contrast with the corresponding results in the nonunique-winner model. To indicate that a problem, such as online-$\mathcal{E}$-UCM, is in the unique-winner model, we write online-$\mathcal{E}$-UCM$_{\text{UW}}$ and ask whether the current manipulator $u$ (assuming support from the manipulators coming after her) can ensure that there exists some $c \in C$ such that $c \geq_\sigma d$ and $c$ is *the unique $\mathcal{E}$ winner* of the election.

## 4. GENERAL RESULTS

THEOREM 3. *(1) For each election system $\mathcal{E}$ whose weighted winner problem can be solved in polynomial time,[2] the problem online-$\mathcal{E}$-WCM is in PSPACE. (2) For each election system $\mathcal{E}$ whose winner problem can be solved in polynomial time, the problem online-$\mathcal{E}$-UCM is in PSPACE. (3) There exists an election system $\mathcal{E}$ with a polynomial-time winner problem such that the problem online-$\mathcal{E}$-UCM is PSPACE-complete. (4) There exists an election system $\mathcal{E}$ with a polynomial-time weighted winner problem such that the problem online-$\mathcal{E}$-WCM is PSPACE-complete.*

The proof of Theorem 3 is deferred to the appendix. The following theorem shows that for bounded numbers of manipulators the complexity crawls up the polynomial hierarchy. The theorem's proof is based on the proof given above, except we need to use the alternating quantifier characterization due to Meyer and Stockmeyer [28] and Stockmeyer [31] for the upper bound and to reduce from the $\Sigma_{2k}^p$-complete problem QBF$_{2k}$ rather than from QBF for the lower bound.

THEOREM 4. *Fix any $k \geq 1$. (1) For each election system $\mathcal{E}$ whose weighted winner problem can be solved in polynomial time, the problem online-$\mathcal{E}$-WCM[k] is in $\Sigma_{2k}^p$. (2) For each election system $\mathcal{E}$ whose winner problem can be solved in polynomial time, the problem online-$\mathcal{E}$-UCM[k] is in $\Sigma_{2k}^p$. (3) There exists an election system $\mathcal{E}$ with a polynomial-time winner problem such that the problem online-$\mathcal{E}$-UCM[k] is $\Sigma_{2k}^p$-complete. (4) There exists an election system $\mathcal{E}$ with a polynomial-time weighted winner problem such that the problem online-$\mathcal{E}$-WCM[k] is $\Sigma_{2k}^p$-complete.*

Note that the (constructive) online manipulation problems considered in Theorems 3 and 4 are about ensuring that the winner set always contains some candidate in the $\sigma$ segment stretching from $d$ up to the top-choice. Now consider "pinpoint" variants of these problems, where we ask whether the distinguished candidate $d$ herself can be guaranteed to be a winner (for nonsequential manipulation, that version indeed is the one commonly studied).

---
[2]We mention in passing here, and henceforward we will not explicitly mention it in the analogous cases, that the claim clearly remains true even when "polynomial time" is replaced by the larger class "polynomial space."

Denote the *pinpoint* variant of, e.g., online-$\mathcal{E}$-UCM[k] by pinpoint-online-$\mathcal{E}$-UCM[k]. Since our hardness proofs in Theorems 3 and 4 make all or no one a winner (and as the upper bounds in these theorems also can be seen to hold for the pinpoint variants), they establish the corresponding completeness results also for the pinpoint cases. We thus have completeness results for PSPACE and $\Sigma_{2k}^p$ for each $k \geq 1$. What about the classes $\Sigma_{2k-1}^p$ and $\Pi_k^p$, for each $k \geq 1$? We can get completeness results for all these classes by defining appropriate variants of online manipulation problems. Let OMP be any of the online manipulation problems considered earlier, including the pinpoint variants mentioned above. Define freeform-OMP to be just as OMP, except we no longer require the distinguished voter $u$ to be part of the manipulative coalition—$u$ can be in or can be out, and the input must specify, for $u$ and all voters after $u$, which ones are the members of the coalition. The question of freeform-OMP is whether it is true that for all actions of the nonmanipulators at or after $u$ (for specificity as to this problem: if $u$ is a nonmanipulator, it will in the input come with a preference order) there will be actions (each taken with full information on cast-before-them votes) of the manipulative coalition members such that their goal of making some candidate $c$ with $c \geq_\sigma d$ (or exactly $d$, in the pinpoint versions) a winner is achieved. Then, whenever Theorem 4 establishes a $\Sigma_{2k}^p$ or $\Sigma_{2k}^p$-completeness result for OMP, we obtain a $\Pi_{2k+1}^p$ or $\Pi_{2k+1}^p$-completeness result for freeform-OMP and for $k = 0$ manipulators we obtain $\Pi_1^p = $ coNP or coNP-completeness results. Similarly, the PSPACE and PSPACE-completeness results for OMP we established in Theorem 3 also can be shown true for freeform-OMP.

On the other hand, if we define a variant of OMP by requiring the final voter to always be a manipulator, the PSPACE and PSPACE-completeness results for OMP from Theorem 3 remain true for this variant; the $\Sigma_{2k}^p$ and $\Sigma_{2k}^p$-completeness results for OMP from Theorem 4 change to $\Sigma_{2k-1}^p$ and $\Sigma_{2k-1}^p$-completeness results for this variant; and the above $\Pi_{2k+1}^p$ and $\Pi_{2k+1}^p$-completeness results for freeform-OMP change to $\Pi_{2k}^p$ and $\Pi_{2k}^p$-completeness results for this variant, $k \geq 1$.

Finally, as an open direction (and related conjecture), we define for each of the previously considered variants of online manipulation problems a *full profile* version. For example, fullprofile-online-$\mathcal{E}$-UCM[k] (for a given election system $\mathcal{E}$) is the function problem that, given an OMS *without* any distinguished candidate, $(C, u, V, \sigma)$, returns a length $\|C\|$ bit-vector that for each candidate $d \in C$ says if the answer to "$(C, u, V, \sigma, d) \in$ online-$\mathcal{E}$-UCM[k]?" is "yes" (1) or "no" (0). The function problem fullprofile-pinpoint-online-$\mathcal{E}$-UCM[k] is defined analogously, except regarding pinpoint-online-$\mathcal{E}$-UCM[k].

It is not hard to prove, as a corollary to Theorem 4, that:

THEOREM 5. *For each election system $\mathcal{E}$ whose winner problem can be solved in polynomial time, (1) the problem fullprofile-online-$\mathcal{E}$-UCM[k] is in $\text{FP}^{\Sigma_{2k}^p[\mathcal{O}(\log n)]}$, the class of functions computable in polynomial time given Turing access to a $\Sigma_{2k}^p$ oracle with $\mathcal{O}(\log n)$ queries allowed on size $n$ inputs; (2) fullprofile-pinpoint-online-$\mathcal{E}$-UCM[k] is in $\text{FP}_{\text{tt}}^{\Sigma_{2k}^p}$, the class of functions computable in polynomial time given truth-table access to a $\Sigma_{2k}^p$ oracle.*

We conjecture that both problems are complete for the corresponding class under metric reductions [26], for suitably defined election systems with polynomial-time winner problems.

If the full profile version of an online manipulation problem can be computed efficiently, we clearly can also easily solve each of the decision problems involved by looking at the corresponding bit of the length $\|C\|$ bit-vector. Conversely, if there is an efficient algorithm for an online manipulation decision problem, we can easily solve its full profile version by running this algorithm for each candidate in turn. Thus, we will state our later results only for online manipulation decision problem.

PROPOSITION 6. *Let* OMP *be any of the online manipulation decision problems defined above. Then* fullprofile-OMP *is in* FP *if and only if* OMP *is in* P.

## 5. RESULTS FOR SPECIFIC NATURAL VOTING SYSTEMS

The results of the previous section show that, simply put, even for election systems with polynomial-time winner problems, online manipulation can be tremendously difficult. But what about *natural* election systems? We will now take a closer look at important natural systems. We will show that online manipulation can be easy for them, depending on which particular problem is considered, and we will also see that the constructive and destructive cases can differ sharply from each other and that it really matters whether we are in the nonunique-winner model or the unique-winner model. Finally, in studying the complexity of online manipulation of veto elections, we obtain (as Theorems 11 and 12) what to the best of our knowledge are the first $P^{NP[1]}$-completeness and $P^{NP}$-completeness results in the field of computational social choice.

THEOREM 7. *(1)* online-plurality-WCM *(and thus also* online-plurality-UCM*) is in* P. *(2)* online-plurality-DWCM *(and thus also* online-plurality-DUCM*) is in* P.

Theorem 7 refers to problems in the nonunique-winner model. By contrast, we now show that online manipulation for weighted plurality voting in the *unique-winner* model is coNP-hard in the *constructive* case and is NP-hard in the *destructive* case. This is perhaps the most dramatic, broad contrast yet between the nonunique-winner model and the unique-winner model, and is the first such contrast involving plurality. The key other NP-hardness versus P result for the nonunique-winner model versus the unique-winner model is due to Faliszewski, Hemaspaandra, and Schnoor [17], but holds only for (standard) weighted manipulation for Copeland$^\alpha$ elections ($0 < \alpha < 1$) with exactly three candidates; for fewer than three both cases there are in P and for more than three both are NP-complete. In contrast, the P results of Theorem 7 hold for all numbers of candidates, and the NP-hardness and coNP-hardness results of Theorem 8 hold whenever there are at least two candidates.

THEOREM 8. *(1)* online-plurality-DWCM$_{UW}$ *is* NP-*hard, even when restricted to only two candidates (and this also holds when restricted to three, four, ... candidates).*

*(2)* online-plurality-WCM$_{UW}$ *is* coNP-*hard, even when restricted to only two candidates (and this also holds when restricted to three, four, ... candidates).*

PROOF. For the first statement, we prove NP-hardness of online-plurality-DWCM$_{UW}$ by a reduction from the NP-complete problem Partition: Given a nonempty sequence $(w_1, w_2, \ldots, w_z)$ of positive integers such that $\sum_{i=1}^{z} w_i = 2W$ for some positive integer $W$, does there exist a set $I \subseteq \{1, 2, \ldots, z\}$ such that $\sum_{i \in I} w_i = W$? Let $m \geq 2$. Given an instance $(w_1, w_2, \ldots, w_z)$ of Partition, construct an instance $(\{c_1, \ldots, c_m\}, u_1, V, c_1 > c_2 > \cdots > c_m, c_1)$ of online-plurality-DWCM$_{UW}$ such that $V$ contains $m + z - 2$ voters $v_1, \ldots, v_{m-2}, u_1, \ldots, u_z$ who vote in that order. For $1 \leq i \leq m-2$, $v_i$ votes for $c_i$ and has weight $(m-1)W - i$, and for $1 \leq i \leq z$, $u_i$ is a manipulator of weight $(m-1)w_i$. If $(w_1, w_2, \ldots, w_z)$ is a yes-instance of Partition, the manipulators can give $(m-1)W$ points to both $c_{m-1}$ and $c_m$, and zero points to the other candidates. So $c_{m-1}$ and $c_m$ are tied for the most points and there is no unique winner. On the other hand, the only way to avoid having a unique winner in our online-plurality-DWCM$_{UW}$ instance is if there is a tie for the most points. The only candidates that can tie are $c_{m-1}$ and $c_m$, since all other pairs of candidates have different scores modulo $m-1$. It is easy to see that $c_{m-1}$ and $c_m$ tie for the most points only if they both get exactly $(m-1)W$ points. It follows that $(w_1, w_2, \ldots, w_z)$ is a yes-instance of Partition.

For the second part, we adapt the above construction to yield a reduction from Partition to the complement of online-plurality-WCM$_{UW}$. Given an instance $(w_1, w_2, \ldots, w_z)$ of Partition, construct an instance $(\{c_1, \ldots, c_m\}, \widehat{u}, V, c_1 > c_2 > \cdots > c_m, c_m)$ of online-plurality-WCM$_{UW}$ such that $V$ contains $m + z - 1$ voters $v_1, \ldots, v_{m-2}, \widehat{u}, u_1, \ldots, u_z$ who vote in that order. For $1 \leq i \leq m-2$, $v_i$ has the same vote and the same weight as above, $\widehat{u}$ is a manipulator of weight 0, and for $1 \leq i \leq z$, $u_i$ has the same weight as above, but in contrast to the case above, $u_i$ is now a nonmanipulator. By the same argument as above, it follows that $(w_1, w_2, \ldots, w_z)$ is a yes-instance of Partition if and only if the nonmanipulators can ensure that there is no unique winner, which in turn is true if and only if the manipulator can not ensure that there is a unique winner. □

THEOREM 9. *For each scoring rule* $\alpha = (\alpha_1, \ldots, \alpha_m)$, online-$\alpha$-WCM *is in* P *if* $\alpha_2 = \alpha_m$ *and is* NP-*hard otherwise.*

THEOREM 10. *For each* $k$, online-$k$-approval-UCM *and* online-$k$-veto-UCM *are in* P.

PROOF. Consider 1-veto. Given an online-1-veto-UCM instance $(C, u, V, \sigma, d)$, the best strategy for the manipulators from $u$ onward (let $n_1$ denote how many of these there are) is to minimize $\max_{c <_\sigma d} score(c)$. Let $n_0$ denote how many nonmanipulators come after $u$. We claim that $(C, u, V, \sigma, d)$ is a yes-instance if and only if $d$ is ranked last in $\sigma$ or there exists a threshold $t$ such that (1) $\sum_{c <_\sigma d}(maxscore(c) \ominus t) \leq n_1$ (so those manipulators can ensure that all candidates ranked $<_\sigma d$ score at most $t$ points), where "$\ominus$" denotes proper subtraction ($x \ominus y = \max(x - y, 0)$) and $maxscore(c)$ is $c$'s score when none of the voters from $u$ onward veto $c$, and

(2) $\sum_{c \geq_\sigma d}(maxscore(c) \ominus (t - 1)) > n_0$ (so those nonmanipulators cannot prevent that some candidate ranked $\geq_\sigma d$ scores at least $t$ points).

For 1-veto under the above approach, in each situation where the remaining manipulators can force success against all actions of the remaining nonmanipulators, $u$ (right then as she moves) can set her *and all future manipulators' actions* so as to force success regardless of the actions of the remaining nonmanipulators. For $k$-approval and $k$-veto, $k \geq 2$, that approach provably cannot work (as will be explained right after this proof); rather, we sometimes need later manipulators' actions to be shaped by intervening nonmanipulators' actions. Still, the following P-time algorithm, which works for all $k$, tells whether success can be forced. As a thought experiment, for each voter $v$ from $u$ onwards in sequence do this: Order the candidates in $\{c \mid c \geq_\sigma d\}$ from most to least current approvals, breaking ties arbitrarily, and postpend the remaining candidates ordered from least to most current approvals. Let $\ell$ be $k$ for $k$-approval and $\|C\| - k$ for $k$-veto. Cast the voter's $\ell$ approvals for the first $\ell$ candidates in this order if $v$ is a manipulator, and otherwise for the last $\ell$ candidates in this order. Success can be forced against perfect play if and only if this P-time process leads to success. □

In the above proof we said that the approach for 1-veto (in which the current manipulator can set her and all future manipulators' actions so as to force success independent of the actions of intervening future nonmanipulators) provably cannot work for $k$-approval and $k$-veto, $k \geq 2$. Why not? Consider an OMS $(C, u, V, \sigma, d)$ with candidate set $C = \{c_1, c_2, \ldots, c_{2k}\}$, $\sigma$ being given by $c_1 >_\sigma c_2 >_\sigma \cdots >_\sigma c_{2k}$, and $d = c_1$. So, $u$'s coalition wants to enforce that $c_1$ is a winner. Suppose that $v_1$ has already cast her vote, now it's $v_2 = u$'s turn, and the order of the future voters is $v_3, v_4, \ldots, v_{2j}$, where all $v_{2i}$, $2 \leq i \leq j$, belong to $u$'s coalition, and all $v_{2i-1}$ do not. Suppose that $v_1$ was approving of the $k$ candidates in $C_1 \subseteq \{c_2, c_3, \ldots, c_{2k}\}$, $\|C_1\| = k$. Then $u$ must approve of the $k$ candidates in $\overline{C_1}$, to ensure that $c_1$ draws level with the candidates in $C_1$ and none of these candidates can gain another point. Next, suppose that nonmanipulator $v_3$ approves of the $k$ candidates in $C_3 \subseteq \{c_2, c_3, \ldots, c_{2k}\}$, $\|C_3\| = k$. Then $v_4$, the next manipulator, must approve of all candidates in $\overline{C_3}$, to ensure that $c_1$ draws level with the candidates in $C_3$ and none of these candidates can gain another point. This process is repeated until the last nonmanipulator, $v_{2j-1}$, approves of the candidates in $C_{2j-1} \subseteq \{c_2, c_3, \ldots, c_{2k}\}$, $\|C_{2j-1}\| = k$, and $v_{2j}$, the final manipulator, is forced to counter this by approving of all candidates in $\overline{C_{2j-1}}$, to ensure that $c_1$ is a winner. This shows that there can be arbitrarily long chains such that the action of each manipulator after $u$ depends on the action of the preceding intervening nonmanipulator.

We now turn to online weighted manipulation for veto when restricted to three candidates. We denote this restriction of online-veto-WCM by online-veto$_{|3}$-WCM.

THEOREM 11. *online-veto$_{|3}$-WCM is $\mathrm{P}^{\mathrm{NP}[1]}$-complete.*

Moving from three to four candidates increases the complexity, namely to $\mathrm{P}^{\mathrm{NP}}$-completeness, and that same bound holds for unlimitedly many candidates. Although this is a strict increase in complexity from $\mathrm{P}^{\mathrm{NP}[1]}$-completeness (unless the polynomial hierarchy collapses [24]), membership in $\mathrm{P}^{\mathrm{NP}}$ still places this problem far below the general PSPACE bound from earlier in this paper. The proof of Theorem 12 is deferred to the appendix. Immediately from Theorems 10 and 12, we have that the full profile variants of online-$k$-veto-UCM and online-$k$-approval-UCM are in FP and that fullprofile-online-veto-WCM is in $\mathrm{FP}^{\mathrm{NP}}$.

THEOREM 12. *online-veto-WCM is $\mathrm{P}^{\mathrm{NP}}$-complete, even when restricted to only four candidates.*

## 6. UNCERTAINTY ABOUT THE ORDER OF FUTURE VOTERS

So far, we have been dealing with cases where the order of future voters was fixed and known. But what happens if the order of future voters itself is unknown? Even here, we can make claims. To model this most naturally, our "magnifying-glass moment" will focus not on one manipulator $u$, but will focus at a moment in time when some voters are still to come (as before, we know who they are and which are manipulators; as before, we have a preference order $\sigma$, and know what votes have been cast so far, and have a distinguished candidate $d$). And the question our problem is asking is: Is it the case that our manipulative coalition can ensure that the winner set will always include $d$ or someone liked more than $d$ with respect to $\sigma$ (i.e., the winner set will have nonempty intersection with $\{c \in C \mid c \geq_\sigma d\}$), *regardless of what order the remaining voters vote in*. We will call this problem the *schedule-robust online manipulation problem*, and will denote it by SR-online-$\mathcal{E}$-UCM. (We will add a "[1,1]" suffix for the restriction of this problem to instances when at most one manipulator and at most one nonmanipulator have not yet voted.) One might think that this problem captures both a $\Sigma_2^p$ and a $\Pi_2^p$ issue, and so would be hard for both classes. However, the requirement of schedule robustness tames the problem (basically what underpins that is simply that exists-forall-predicate implies forall-exists-predicate), bringing it into $\Sigma_2^p$. Further, we can prove, by explicit construction of such a system, that for some simple election systems this problem is complete for $\Sigma_2^p$.

THEOREM 13. *(1) For each election system $\mathcal{E}$ whose winner problem is in P, SR-online-$\mathcal{E}$-UCM is in $\Sigma_2^p$. (2) There exists an election system $\mathcal{E}$, whose winner problem is in P, such that the problem SR-online-$\mathcal{E}$-UCM (indeed, even SR-online-$\mathcal{E}$-UCM[1, 1]) is $\Sigma_2^p$-complete.*

## 7. CONCLUSIONS AND OPEN QUESTIONS

We introduced a novel framework for online manipulation in sequential voting, and showed that manipulation there can be tremendously complex even for systems with simple winner problems. We also showed that among the most important election systems, some have efficient online manipulation algorithms but others (unless P = NP) do not. It will be important to, complementing our work, conduct typical-case complexity studies (although we mention in passing that unless the polynomial hierarchy collapses, no heuristic algorithm for any NP-hard problem can have a subexponential error rate, see the discussion in the survey [23]). We have extended the scope of our investigation by studying online control [22, 21] and will also study online bribery.

## 8. ACKNOWLEDGMENTS

We thank the anonymous reviewers for helpful comments. This work was supported in part by NSF grants CCF-{0915792, 1101452, 1101479}, ARC grant DP110101792, DFG grant RO-1202/15-1, SFF grant "Cooperative Norm-setting" of HHU Düsseldorf, Friedrich Wilhelm Bessel Research Awards to Edith Hemaspaandra and Lane A. Hemaspaandra from the Alexander von Humboldt Foundation, and a DAAD grant for a PPP project in the PROCOPE program.

## APPENDIX. DEFERRED PROOFS

We provide here deferred proofs of two of our results that were not proven in the paper's body. Most other proofs not in the body can be found in the technical report version [20].

**Proof of Theorem 3.** The proof of the first statement (which is analogous to the proof of the first statement in Theorem 4) follows from the easy fact that online-$\mathcal{E}$-WCM can be solved by an alternating Turing machine in polynomial time, and thus, due to the characterization of Chandra, Kozen, and Stockmeyer [9], by a deterministic Turing machine in polynomial space. The proof of the second case is analogous.

We construct an election system $\mathcal{E}$ establishing the third statement. Let $(C, u, V, \sigma, d)$ be a given input. $\mathcal{E}$ will look at the lexicographically least candidate name in $C$. Let $c$ represent that name string in some fixed, natural encoding. $\mathcal{E}$ will check if $c$ represents a *tiered* boolean formula, by which we mean one whose variable names are all of the form $x_{i,j}$ (which really means a direct encoding of a string, such as "$x_{4,9}$"); the $i, j$ fields must all be positive integers. If $c$ does not represent such a tiered formula, everyone loses on that input. Otherwise (i.e., if $c$ represents a tiered formula), let *width* be the maximum $j$ occurring as the second subscript in any variable name ($x_{i,j}$) in $c$, and let *blocks* be the maximum $i$ occurring as the first subscript in any variable name in $c$. If there are fewer than *blocks* voters in $V$, everyone loses. Otherwise, if there are fewer than $1 + 2 \cdot width$ candidates in $C$, everyone loses (this is so that each vote will involve enough candidates that it can be used to set all the variables in one block). Otherwise, if there exists some $i$, $1 \leq i \leq$ *blocks*, such that for no $j$ does the variable $x_{i,j}$ occur in $c$, then everyone loses. Otherwise, order the voters from the lexicographically least to the lexicographically greatest voter name. If distinct voters are allowed to have the same name string (e.g., John Smith), we break ties by sorting according to the associated preference orders within each group of tied voters (second-order ties are no problem, as those votes are identical, so any order will have the same effect). Now, the first voter in this order will assign truth values to all variables $x_{1,*}$, the second voter in this order will assign truth values to all variables $x_{2,*}$, and so on up to the *blocks*th voter, who will assign truth values to all variables $x_{blocks,*}$.

How do we get those assignments from these votes? Consider a vote whose total order over $C$ is $\sigma'$ (and recall that $\|C\| \geq 1 + 2 \cdot width$). Remove $c$ from $\sigma'$, yielding $\sigma''$. Let $c_1 <_{\sigma''} c_2 <_{\sigma''} \cdots <_{\sigma''} c_{2 \cdot width}$ be the $2 \cdot width$ least preferred candidates in $\sigma''$. We build a vector in $\{0,1\}^{width}$ as follows: The $\ell$th bit of the vector is 0 if the string that names $c_{1+2(\ell-1)}$ is lexicographically less than the string that names $c_{2\ell}$, and this bit is 1 otherwise.

Let $b_i$ denote the vector thus built from the $i$th vote (in the above ordering), $1 \leq i \leq$ *blocks*. Now, for each variable $x_{i,j}$ occurring in $c$, assign to it the value of the $j$th bit of $b_i$, where 0 represents *false* and 1 represents *true*. We have now assigned all variables of $c$, so $c$ evaluates to either *true* or *false*. If $c$ evaluates to *true*, everyone wins, otherwise everyone loses. This completes the specification of the election system $\mathcal{E}$. $\mathcal{E}$ has a polynomial-time winner problem, as any boolean formula, given an assignment to all its variables, can easily be evaluated in polynomial time.

To show PSPACE-hardness, we $\leq_m^p$-reduce the PSPACE-complete problem QBF to the problem online-$\mathcal{E}$-UCM. Let $y$ be an instance of QBF. We transform $y$ into an instance of the form $(\exists x_{1,1}, x_{1,2}, \ldots, x_{1,k_1})(\forall x_{2,1}, x_{2,2}, \ldots, x_{2,k_2}) \cdots (Q_\ell x_{\ell,1}, x_{\ell,2}, \ldots, x_{\ell,k_\ell}) [\Phi(x_{1,1}, x_{1,2}, \ldots, x_{1,k_1}, x_{2,1}, x_{2,2}, \ldots, x_{2,k_2}, \ldots, x_{\ell,1}, x_{\ell,2}, \ldots, x_{\ell,k_\ell})]$ in polynomial time, where $Q_\ell = \exists$ if $\ell$ is odd and $Q_\ell = \forall$ if $\ell$ is even, the $x_{i,j}$ are boolean variables, $\Phi$ is a boolean formula, and for each $i$, $1 \leq i \leq \ell$, $\Phi$ contains at least one variable of the form $x_{i,*}$. This quantified boolean formula is $\leq_m^p$-reduced to an instance $(C, u, V, \sigma, c)$ of online-$\mathcal{E}$-UCM as follows:

1. $C$ contains a candidate whose name, $c$, encodes $\Phi$, and in addition $C$ contains $2 \cdot \max(k_1, \ldots, k_\ell)$ other candidates, all with names lexicographically greater than $c$—for specificity, let us say their names are the $2 \cdot \max(k_1, \ldots, k_\ell)$ strings that immediately follow $c$ in lexicographic order.

2. $V$ contains $\ell$ voters, $1, 2, \ldots, \ell$, who vote in that order, where $u = 1$ is the distinguished voter and all odd voters belong to $u$'s manipulative coalition and all even voters do not. The voter names will be lexicographically ordered by their number, 1 is least and $\ell$ is greatest.

3. The manipulators' preference order $\sigma$ is to like candidates in the opposite of their lexicographic order. In particular, $c$ is the coalition's most preferred candidate.

This is a polynomial-time reduction. It follows immediately from this construction and the definition of $\mathcal{E}$ that $y$ is in QBF if and only if $(C, u, V, \sigma, c)$ is in online-$\mathcal{E}$-UCM.

To prove the last statement, simply let $\mathcal{E}$ be the election system that ignores the weights of the voters and then works exactly as the previous election system. ❑ Theorem 3

**Proof of Theorem 12.** We first show that online-veto-WCM is in $P^{NP}$. The proof is reminiscent of the proof for 1-veto in Theorem 10. Let $(C, u, V, \sigma, d)$ be a given instance of online-veto-WCM with $C = \{c_1, c_2, \ldots, c_m\}$ and $c_1 >_\sigma c_2 >_\sigma \cdots >_\sigma c_m$. Suppose $d = c_i$. Our $P^{NP}$ algorithm proceeds as follows:

1. Compute the minimal threshold $t_1$ such that there exists a partition $(A_{i+1}, \ldots, A_m)$ of the weights of the manipulators from $u$ onward such that for each $j$, $i+1 \leq j \leq m$, $maxscore(c_j) - \sum A_j \leq t_1$, where $maxscore(c_j)$ is $c_j$'s score when none of the voters from $u$ onward veto $c$. That is, by having manipulators from $u$ onward with weights in $A_j$ veto $c_j$, the manipulators from $u$ onward can ensure that none of the candidates they dislike more than $d$ exceeds a score of $t_1$.

2. Compute the minimal threshold $t_2$ such that there exists a partition $(A_1, \ldots, A_i)$ of the weights of the non-manipulators after $u$ such that for each $j$, $1 \leq j \leq i$, $maxscore(c_j) - \sum A_j \leq t_2$. That is, if the nonmanipulators after $u$ with weights in $A_j$ veto $c_j$, none of the candidates that the manipulators like as least as much as $d$ exceeds a score of $t_2$.

3. Accept if and only if $t_1 \leq t_2$.

Note that the first two steps of the algorithm can both be done in $\text{FP}^{\text{NP}}$ by using an NP oracle that checks whether there exists a partition of the specified kind.

It remains to show that online-veto$_{|4}$-WCM is $\text{P}^{\text{NP}}$-hard. We will reduce from the standard $\text{P}^{\text{NP}}$-complete problem MAXSATASG$_=$, which is the set of pairs of 3cnf formulas[3] that have the same maximal satisfying assignment [34]. To be precise, we will assume that our propositional variables are $x_1, x_2, \ldots$. If $x_n$ is the largest propositional variable occurring in $\phi$, we often write $\phi(x_1, \ldots, x_n)$ to make that explicit. An assignment for $\phi(x_1, \ldots, x_n)$ is an $n$-bit string $\alpha$ such that $\alpha_i$ gives the assignment for variable $x_i$. We will sometimes identify $\alpha$ with the binary integer it represents. For $\phi$ a formula, $maxsatasg(\phi)$ is the lexicographically largest satisfying assignment for $\phi$. If $\phi$ is not satisfiable, $maxsatasg(\phi)$ is not defined. And we define MAXSATASG$_=$ as the set of pairs of 3cnf formulas $(\phi(x_1, \ldots, x_n), \psi(x_1, \ldots, x_n))$ such that $\phi$ and $\psi$ are satisfiable 3cnf formulas, and $maxsatasg(\phi) = maxsatasg(\psi)$.

The OMS that we will construct will have four candidates, $a >_\sigma b >_\sigma c >_\sigma d$, and the distinguished candidate will be $b$. Looking at the $\text{P}^{\text{NP}}$ algorithm above, we can see that determining whether the OMS can be manipulated basically amounts to determining whether the nonmanipulator weights have a "better" partition than the manipulator weights.

So, we will associate formulas with multisets of positive integers, and their satisfying assignments with subset sums. This already happens in the standard reduction from 3SAT to SubsetSum. However, we also want larger satisfying assignments to correspond to "better" subset sums. In order to do this, we use Wagner's variation of the 3SAT to SubsetSum reduction [34]. Wagner uses this reduction to prove that determining whether the largest subset sum up to a certain bound is odd is a $\text{P}^{\text{NP}}$-hard problem.

LEMMA 14. *Let $\phi(x_1, \ldots, x_n)$ be a 3cnf formula. Wagner's reduction maps this formula to an instance $(k_1, \ldots, k_t, L)$ of SubsetSum with the following properties:*

1. *For all assignments $\alpha$, $\phi[\alpha]$ if and only if there exists a subset of $k_1, \ldots, k_t$ that sums to $L + \alpha$.*

2. *For all $K$ such that $2^n \leq K \leq 2(2^n - 1)$, no subset of $k_1, \ldots, k_t$ sums to $L + K$.*

**Proof of Lemma 14.** The first claim is immediate from the proof of Theorem 8.1(3) from [34]. For the second claim, note that $L + K \leq L + 2(2^n - 1) < L + 6^n$. In Wagner's construction, $L = \underbrace{3 \cdots 3}_{m}\underbrace{1 \cdots 1}_{n}\underbrace{0 \cdots 0}_{n}$ in base 6, where $m$ is the number of clauses in $\phi$. So, $(L + K)$'s representation in base 6 is $\underbrace{3 \cdots 3}_{m}\underbrace{1 \cdots 1}_{n}$ followed by $n$ digits. It is easy to see from Wagner's construction that the subset sums of

---
[3]We denote a formula in conjunctive normal form by *cnf formula*, and a *3cnf formula* is a cnf formula with exactly three literals per clause.

this form that can be realized are exactly $L + \beta$, where $\beta$ is a satisfying assignment of $\phi$. Since $K \geq 2^n$, $K$ is not even an assignment, and thus no subset of $k_1, \ldots, k_t$ sums to $L + K$. ❑ Lemma 14

Let $\phi(x_1, \ldots, x_n)$ and $\psi(x_1, \ldots, x_n)$ be 3cnf formulas, and consider instance $(\phi, \psi)$ of MAXSATASG$_=$. Without loss of generality, we assume that $x_1$ does not actually occur in $\phi$ or $\psi$. We will define an OMS $(C, u, V, \sigma, b)$ with $C = \{a, b, c, d\}$ and $\sigma = a > b > c > d$ such that $(\phi, \psi) \in$ MAXSATASG$_=$ if and only if $(C, u, V, \sigma, b)$ is a positive instance of online-veto-WCM. Note that MAXSATASG$_=$ corresponds to optimal solutions being equal, while online-veto-WCM corresponds to one optimal solution being at least as good as the other. We will first modify the formulas such that we also look at the optimal solution for one formula being at least as good as the optimal solution for the other. The following is immediate.

CLAIM 15. *$(\phi, \psi) \in$ MAXSATASG$_=$ if and only if $\phi \wedge \psi$ is satisfiable and $maxsatasg(\phi \wedge \psi) \geq maxsatasg(\phi \vee \psi)$.*

It will also be very useful if one of the formulas is always satisfiable. We can easily ensure this by adding an extra variable that will correspond to the highest order bit of the satisfying assignment. Recall that $x_1$ does not occur in $\phi$ or $\psi$.

CLAIM 16. *$(\phi, \psi) \in$ MAXSATASG$_=$ if and only if $\phi \wedge \psi \wedge x_1$ is satisfiable and*

$maxsatasg(\phi \wedge \psi \wedge x_1) \geq maxsatasg(\phi \vee \psi \vee \overline{x_1}).$

Now we would like to apply the reduction from Lemma 14 on $\phi \wedge \psi \wedge x_1$ and $\phi \vee \psi \vee \overline{x_1}$. But wait! This reduction is defined for 3cnf formulas, and $\phi \vee \psi \vee \overline{x_1}$ is not in 3cnf. Since $\phi$ and $\psi$ are in 3cnf, it is easy to convert $\phi \vee \psi \vee \overline{x_1}$ into cnf in polynomial time. Let $g$ be the standard reduction from CNF-SAT to 3SAT. We can rename the variables such that $g$ has the following property: For $\xi(x_1, \ldots, x_n)$ a cnf formula, $g(\xi)(x_1, \ldots, x_n, x_{n+1}, \ldots, x_{\hat{n}})$ is a 3cnf formula such that $\hat{n} > n$ and such that for all assignments $\alpha \in \{0,1\}^n$, $\xi[\alpha]$ if and only if there exists an assignment $\beta \in \{0,1\}^{\hat{n}-n}$ such that $g(\xi)[\alpha\beta]$.

Let $\widehat{\psi}(x_1, \ldots, x_{\hat{n}}) = g(\phi \vee \psi \vee \overline{x_1})$. Let $\widehat{\phi}(x_1, \ldots, x_{\hat{n}}) = \phi \wedge \psi \wedge (x_1 \vee x_1 \vee x_1) \wedge (x_{\hat{n}} \vee x_{\hat{n}} \vee \overline{x_{\hat{n}}})$.

CLAIM 17.  
- $\widehat{\phi}$ and $\widehat{\psi}$ are in 3cnf and $\widehat{\psi}$ is satisfiable.
- $(\phi, \psi) \in$ MAXSATASG$_=$ if and only if $\widehat{\phi}$ is satisfiable and $maxsatasg(\widehat{\phi}) \geq maxsatasg(\widehat{\psi})$.

**Proof of Claim 17.** From the previous claim we know that if $(\phi, \psi) \in$ MAXSATASG$_=$, then $\phi \wedge \psi \wedge x_1$ is satisfiable and thus $\widehat{\phi}$ is satisfiable. Also from the previous claim, if $(\phi, \psi) \in$ MAXSATASG$_=$, then $maxsatasg(\phi \wedge \psi \wedge x_1) \geq maxsatasg(\phi \vee \psi \vee \overline{x_1})$. Let $\alpha$ be the maximal satisfying assignment of $\phi \wedge \psi \wedge x_1$. Then $\alpha 1^{\hat{n}-n}$ is the maximal satisfying assignment of $\widehat{\phi}$. Let $\alpha'$ be the maximal satisfying assignment of $\phi \vee \psi \vee \overline{x_1}$. Then $\alpha'\beta$ is the maximal satisfying assignment of $\widehat{\psi}$ for some $\beta$. Since $\alpha \geq \alpha'$, it follows that $\alpha 1^{\hat{n}-n} \geq \alpha'\beta$.

For the converse, suppose that $\widehat{\phi}$ is satisfiable and $maxsatasg(\widehat{\phi}) \geq maxsatasg(\widehat{\psi})$. Let $\gamma$ be the maximal satisfying assignment of $\widehat{\phi}$ and let $\gamma'$ be the maximal satisfying assignment of $\widehat{\psi}$. Then the length-$n$ prefix of $\gamma$ is the maximal satisfying assignment of $\phi \wedge \psi \wedge x_1$ and the length-$n$ prefix of $\gamma'$ is the maximal satisfying assignment of $\phi \vee \psi \vee \overline{x_1}$. Since $\gamma \geq \gamma'$, the $n$-bit prefix of $\gamma$ is greater than or equal to the $n$-bit prefix of $\gamma'$. ☐ Claim 17

We now apply Wagner's reduction from Lemma 14 to $\widehat{\phi}$ and $\widehat{\psi}$. Let $k_1, \ldots, k_t, L$ be the output of Wagner's reduction on $\widehat{\phi}$ and let $k'_1, \ldots, k'_{t'}, L'$ be the output of Wagner's reduction on $\widehat{\psi}$.

As mentioned previously, we will define an OMS $(C, u, V, \sigma, b)$ with $C = \{a, b, c, d\}$ and $\sigma = a > b > c > d$ such that $(\phi, \psi) \in \text{MAXSATASG}_=$ if and only if $(C, u, V, \sigma, b)$ is a positive instance of online-veto-WCM. Because we are looking at veto, when determining the outcome of an election, it is easiest to simply count the number of vetoes for each candidate. Winners have the fewest vetoes. For $\hat{c}$ a candidate, we will denote the total weight of the voters that veto $\hat{c}$ by $vetoes(\hat{c})$.

There are four voters in $V_{<u}$: one voter of weight $L$ vetoing $a$, one voter of weight $L + 2L' + 2(2^{\hat{n}} - 1) - \sum k'_i$ vetoing $b$, one voter of weight $L'$ vetoing $c$, and one voter of weight $L' + 2L + 2(2^{\hat{n}} - 1) - \sum k_i$ vetoing $d$. Let $u = u_1$. $V_{u<}$ consists of $t - 1$ further manipulators $u_2, \ldots, u_t$ followed by $t'$ nonmanipulators $u'_1, \ldots, u'_{t'}$. The weight of manipulator $u_i$ is $k_i$ and the weight of nonmanipulator $u'_i$ is $k'_i$.

It remains to show that the reduction is correct. First suppose that $(\phi, \psi)$ is in MAXSATASG$_=$. By Claim 17, this implies that $\widehat{\phi}$ and $\widehat{\psi}$ are satisfiable 3cnf formulas such that $maxsatasg(\widehat{\phi}) \geq maxsatasg(\widehat{\psi})$. Let $\alpha = maxsatasg(\widehat{\phi})$. We know from Lemma 14 that there exists a subset of $k_1, \ldots, k_t$ that sums to $L + \alpha$. The manipulators corresponding to this subset will veto $c$, so that $c$ receives $L + \alpha$ vetoes from the manipulators. The remaining manipulators will veto $d$, i.e., $d$ receives $(\sum k_i) - L - \alpha$ vetoes from the manipulators. After the manipulators have voted, $vetoes(a) = L$, $vetoes(b) = L + 2L' + 2(2^{\hat{n}} - 1) - \sum k'_i$, $vetoes(c) = L' + L + \alpha$, and $vetoes(d) = L' + L + 2(2^{\hat{n}} - 1) - \alpha$. Since $\alpha \leq 2^{\hat{n}} - 1$, $vetoes(c) \leq vetoes(d)$. We will show that no matter how the nonmanipulators vote, $a$ or $b$ is a winner. Suppose for a contradiction that after the nonmanipulators have voted, $vetoes(a) > vetoes(c)$ and $vetoes(b) > vetoes(c)$. If that were to happen, there would be a subset of $k'_1, \ldots, k'_{t'}$ summing to $K$ such that $L + K = vetoes(a) > vetoes(c) = L + L' + \alpha$ and $L + 2L' + 2(2^{\hat{n}} - 1) - K = vetoes(b) > vetoes(c) = L + L' + \alpha$. It follows that $\alpha < K - L' < 2(2^{\hat{n}} - 1)$ and there exists a subset of $k'_1, \ldots, k'_{t'}$ that sums to $L' + (K - L')$. It follows from Lemma 14 that $K - L'$ is a satisfying assignment for $\widehat{\psi}$. But that contradicts the assumption that $maxsatasg(\widehat{\phi}) \geq maxsatasg(\widehat{\psi})$.

The proof of the converse is very similar. Suppose that $(\phi, \psi) \notin$ MAXSATASG$_=$. By Claim 17, $\widehat{\psi}$ is satisfiable. Let $\alpha = maxsatasg(\widehat{\psi})$. By Claim 17, either $\widehat{\phi}$ is not satisfiable or $maxsatasg(\widehat{\phi}) < \alpha$. Suppose the manipulators vote such that $c$ receives $K$ vetoes from some of them. Without loss of generality, assume all other manipulators veto $d$, so that $d$ receives $(\sum k_i) - K$ vetoes from the manipulators. We know from Lemma 14 that there exists a subset of $k'_1, \ldots, k'_{t'}$ that sums to $L' + \alpha$. After the manipulators have voted, the nonmanipulators will vote such that $a$ receives $L' + \alpha$ vetoes from the nonmanipulators and the remaining nonmanipulators will veto $b$, i.e., $b$ receives $(\sum k'_i) - L' - \alpha$ vetoes from the nonmanipulators. So, $vetoes(a) = L + L' + \alpha$, $vetoes(b) = L + L' + 2(2^{\hat{n}} - 1) - \alpha$, $vetoes(c) = L' + K$, and $vetoes(d) = L' + 2L + 2(2^{\hat{n}} - 1) - K$. We will show that neither $a$ nor $b$ is a winner. Since $\alpha \leq 2^{\hat{n}} - 1$, $vetoes(a) \leq vetoes(b)$. So it suffices to show that $a$ is not a winner. If $a$ were a winner, $vetoes(a) \leq vetoes(c)$ and $vetoes(a) \leq vetoes(d)$. This implies that $\alpha \leq K - L \leq 2(2^{\hat{n}} - 1)$. It follows from Lemma 14 that $K - L$ is a satisfying assignment for $\widehat{\phi}$. But that contradicts the assumption that either $\widehat{\phi}$ is not satisfiable or $maxsatasg(\widehat{\phi}) < \alpha$. ☐ Theorem 12